\UseRawInputEncoding

\documentclass[conference]{IEEEtran}
 \usepackage[pdftex]{graphicx}
  \DeclareGraphicsExtensions{.pdf,.jpeg,.png}
\usepackage{algorithm}
\usepackage{algpseudocode}

\usepackage{etoolbox}
\usepackage{tabu}
\usepackage{setspace}
\usepackage{makecell}
\usepackage{epsfig}
\usepackage[usenames, dvipsnames]{color}
\usepackage[roman]{parnotes}

\usepackage{amsthm,amsmath,amssymb}
\usepackage{amsmath}
\usepackage{mathtools}
\usepackage{breqn}
\usepackage{amsfonts}
\usepackage{amssymb}
\usepackage{amsthm}
\usepackage{mathrsfs}
\usepackage{xparse}
\usepackage{multirow}
\usepackage{chngcntr}
\usepackage{dblfloatfix}    
\usepackage{soul} 
\usepackage{cite}
\usepackage{bm}
\usepackage{comment}
\usepackage{footnote}
\usepackage{subcaption}
\usepackage{algorithmicx}
\usepackage{algpseudocode}
\usepackage[font={small}]{caption}

\algdef{SE}[DOWHILE]{Do}{doWhile}{\algorithmicdo}[1]{\algorithmicwhile\ #1}%

\makeatletter
\newenvironment{breakablealgorithm}
  {
   \begin{center}
     \refstepcounter{algorithm}
     \hrule height.8pt depth0pt \kern2pt
     \renewcommand{\caption}[2][\relax]{
       {\raggedright\textbf{\fname@algorithm~\thealgorithm} ##2\par}%
       \ifx\relax##1\relax 
         \addcontentsline{loa}{algorithm}{\protect\numberline{\thealgorithm}##2}%
       \else 
         \addcontentsline{loa}{algorithm}{\protect\numberline{\thealgorithm}##1}%
       \fi
       \kern2pt\hrule\kern2pt
     }
  }{
     \kern2pt\hrule\relax
   \end{center}
  }
\makeatother

\makeatletter
\newcommand*{\algrule}[1][\algorithmicindent]{%
  \makebox[#1][l]{%
    \hspace*{.2em}
    \vrule height .75\baselineskip depth .25\baselineskip
  }
}

\newcount\ALG@printindent@tempcnta
\def\ALG@printindent{%
    \ifnum \theALG@nested>0
    \ifx\ALG@text\ALG@x@notext
    \else
    \unskip
    \ALG@printindent@tempcnta=1
    \loop
    \algrule[\csname ALG@ind@\the\ALG@printindent@tempcnta\endcsname]%
    \advance \ALG@printindent@tempcnta 1
    \ifnum \ALG@printindent@tempcnta<\numexpr\theALG@nested+1\relax
    \repeat
    \fi
    \fi
}
\patchcmd{\ALG@doentity}{\noindent\hskip\ALG@tlm}{\ALG@printindent}{}{\errmessage{failed to patch}}
\patchcmd{\ALG@doentity}{\item[]\nointerlineskip}{}{}{} 
\makeatother

\DeclareMathOperator{\tr}{tr}
\DeclareMathOperator{\diag}{diag}

\newcommand{\argminF}{\mathop{\mathrm{argmin}}\limits}
\allowdisplaybreaks

\begin{document}
%
\title{Wireless communications with user equipment mounted Reconfigurable Intelligent Surfaces}
\author{\IEEEauthorblockN{I. Zakir Ahmed  and Hamid Sadjadpour\\}
\IEEEauthorblockA{Department of Electrical and Computer Engineering\\
University of California, Santa Cruz\\
}}


%

\makeatletter
\def\ps@IEEEtitlepagestyle{%
  \def\@oddhead{\mycopyrightnotice}%
  \def\@oddfoot{\hbox{}\@IEEEheaderstyle\leftmark\hfil\thepage}\relax
  \def\@evenhead{\@IEEEheaderstyle\thepage\hfil\leftmark\hbox{}}\relax
  \def\@evenfoot{}%
}
\def\mycopyrightnotice{%
  \begin{minipage}{\textwidth}
  \scriptsize
Copyright © 2024 IEEE. Personal use of this material is permitted. Permission from IEEE must be obtained for all other uses, in any current or future media, including reprinting/republishing this material for advertising or promotional purposes, creating new collective works, for resale or redistribution to servers or lists, or reuse of any copyrighted component of this work in other works by sending a request to pubs-permissions@ieee.org. Accepted for publication in 2024 Asilomar Conference on Signals, Systems, and Computers.
  \end{minipage}
}
\maketitle


\setlength{\columnsep}{0.21 in}

\maketitle

\begin{abstract}
In traditional Reconfigurable Intelligent Surfaces (RIS) systems, the RIS is mounted on stationary structures like buildings, walls, or posts. They have shown promising results in enhancing the performance of wireless systems like capacity and MSE in poor channel conditions. The traditional RIS is a monolithic structure containing a large number of reflecting elements (passive or active). In this paper, we propose the idea of mounting a small number of RIS elements (usually between 2 to 4 ) on user equipment (UEs) like mobile phones, laptops, and tablets, to name a few. A joint coordinated optimization of phase shifts of all the passive RIS elements on the participating UEs is envisioned to enhance the performance of wireless communication between an intended transmitter and receiver in the MSE sense. Given that the RIS elements are mounted on the UEs, the challenging channel estimation problem with RIS is significantly simplified. For the case when there is a line-of-sight (LOS) channel and with a large number of participating RIS-mounted UEs, the LOS is converted into a multipath-rich-scattering channel even for millimeter wave and Terahertz operating ranges that enable higher spatial multiplexing gains, thereby significantly improving the MSE performance compared to traditional RIS channels. We support the above claims using simulations.
\end{abstract}
%

%

%
\IEEEpeerreviewmaketitle

\section{Introduction}
One of the goals of the 6G standardization effort is to add about $10^6$ million users per square kilometer (sq.km) as compared to one million users per sq.km with 5G \cite{Twelve6G}. This extreme densification goal is expected with all the users experiencing peak data rates of $1$ Tbps, with reliability of 99.99999\%, and air-interface latency of $0.1$ ms \cite{tong_zhu_2021, Twelve6G}. A reconfigurable intelligent surface (RIS)-assisted massive Multiple-Input Multiple-Output (MaMIMO) wireless link has shown immense potential in achieving the 6G goals \cite{6Gsurvey, Zakir11}. The RIS manipulates the wireless channel of interest between the intended transmitter and receiver to improve the throughput, mean-squared-error (MSE) performance, or energy efficiency of the wireless links that may experience obstruction, shadowing, and complex fading scenarios. In addition, the RIS is shown to enable coverage extension, particularly with the millimeter waves (mmWave) and terahertz (THz) communication links due to the unfavorable free-space omnidirectional path loss in these frequency bands \cite{AppIRS,6GIrs,Zakir11}.\\
Traditional RIS-assisted communication consists of a single or multiple RIS (multi-RIS) with a large number of reflecting elements. They are mounted on buildings or walls to enhance the performance of the wireless link between the transmitter and receiver of interest \cite{Zakir11}. In this work, we take advantage of the increased number of UEs available in future wireless networks and propose a RIS system with very few elements mounted on a large number of distributed user equipment (UE) that collectively cooperate to enhance the performance of the wireless link between a transmitter and receiver of interest equipped with MaMIMO. 

\subsection{Previous works}
Distributed RISs (DRIS) have invoked a lot of interest and many research articles have appeared in recent years. The studies indicate that the DRISs have shown promising trends that improve the coverage, ergodic capacity, and outage probability when compared to the single RIS systems. Additionally, the DRIS framework aids sensing applications, which is one of the goals of 6G standards- to integrate communication with sensing \cite{Twelve6G}. However, all the DRIS architectures or multi-RIS based optimization methods proposed in the literature focus on using the traditional RIS structures with large number of reflecting elements that are mounted on buildings or walls \cite{Dris1, Dris2, Dris3, Dris4}. 

\subsection{Our contributions}
\textit{(i)} We propose a UE-mounted-RIS-assisted (UE-RIS) cooperative and distributed communication framework for the next-generation wireless network (NGWN). We consider a large number of cooperating UEs equipped with passive RIS comprising a small number of reflecting elements (typically between $2-4$). The UE-RIS leverages the dense network of UEs envisioned in the NGWN. Importantly, the UE-RIS framework can potentially ease the burden of network planning, which is needed to position multiple static RISs in the network by offloading a portion of it as a dynamic resource allocation problem.\\
\textit{(ii)} We show that the UE-RIS framework enables a much simpler channel estimation scheme compared to the traditional RIS system due to the placement of the RIS elements on an active computing and communicating device (UE).\\
\textit{(iii)} We device an alternating optimization (AO) algorithm to jointly optimize RIS phase shifts and hybrid precoder/combiner. An Information-directed-branch-and-prune (IDBP) algorithm proposed in \cite{Zakir11} is used to optimize the RIS phase shifts within the AO called AO-IDBP. The IDBP algorithm guarantees near-optimal MSE performance.\\

\indent \textit{Notations :}
The column vectors are represented as boldface small letters and matrices as boldface uppercase letters. The primary diagonal of a matrix is denoted as ${\rm diag}(\cdot)$, and all expectations $E[\cdot]$ are over the random variable $\bold{n}$, which is an AWGN vector. The multivariate normal distribution with mean $\boldsymbol{\mu}$ and covariance $\boldsymbol{\varphi}$ is denoted as $\mathcal{N}(\boldsymbol{\mu},\boldsymbol{\varphi})$ and $\mathcal{CN}(\bold{0},{\boldsymbol{\varphi}})$ denotes a multivariate circularly-symmetric Gaussian distribution. The trace of a matrix $\bold{A}$ is shown as $\tr{(\bold{A})}$ and $\bold{I}_N$ represents a $N \times N$ identity matrix. The Frobenius norm of matrix $\bold{A}$ is indicated as $\lVert \bold{A} \rVert_F$. The superscripts T and H denote transpose and Hermitian transpose, respectively.

\section{System Model}\label{sigmod}
\begin{figure}[h!]
\centering
\includegraphics[scale=0.18]{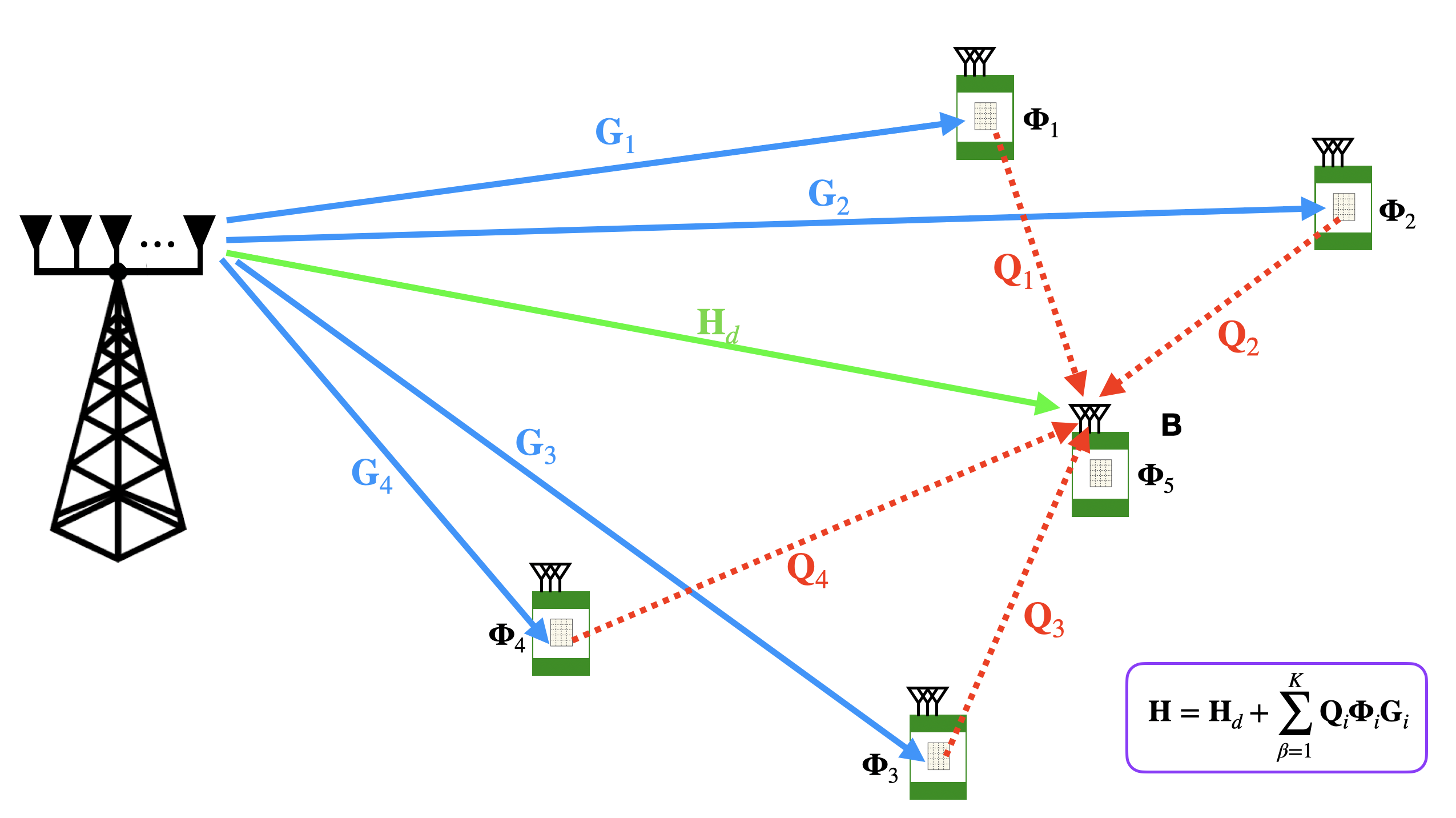}
\caption{\small UE-RIS in cellular network}
\label{fig1}
\end{figure}

A communication system comprising of a transmitter (TX) usually a basestation or an access point, a receiver (RX) usually a UE, and a set of other $N_d$ UEs willing to cooperate to enhance the link  between TX and RX is considered. The TX and RX are assumed to be equipped with MaMIMO antennas, hybrid precoder, and combiner. Each of the $N_d$ UEs are assumed to be equipped with $M$ element RIS such that $M \le 4$. The elements of the RIS are assumed to be placed on a uniform linear strip with the separation between the elements being integer multiples of half wavelength $(d=k \frac{\lambda}{2})$. An example use-case scenarios is illustrated in Fig. \ref{fig1}. This proposed signal model can be extended to other use-case scenarios both within indoor and outdoor wireless cellular or vehicular networks.\\
\indent We denote ${\bold{F}_D}$ and ${\bold{F}_A}$ to be the digital and analog precoders, respectively. We represent ${\bold{W}_D^H}$ and ${\bold{W}_A^H}$ to be the digital and analog combiners, respectively. The vector $\bold{x}$ is an $N \times1$ transmitted signal vector whose average power is unity. Let $N_{rt}$ and $N_{rs}$ denote the number of RF Chains at the transmitter and receiver, respectively. Also, $N_t$ and $N_r$ represent the number of transmit and receive antennas, respectively. The effective channel $\bold{H}$ is a $N_r \times N_t$ matrix at the intended receiver and will be a combination of the RIS reflected signals from $N_d$ cooperating UEs and a line-of-sight (LOS) channel $\bold{H}_d$. The channel model of such a communication system can be written as
\begin{equation}\label{eq_1}
\begin{split}
\bold{H} = \bold{H}_d + \bold{H}_{ris},\text{ such that }\bold{H}_{ris} = \sum_{i=1}^{N_d} \bold{Q}_i\bold{\Phi}_i\bold{G}_i.
\end{split}
\end{equation}
The term $\bold{G}_i \in \mathbb{C}^{M \times N_t}$ is the transimtter-to-RIS (TX-RIS$_i$) channel, $\bold{Q}_i \in \mathbb{C}^{N_r \times M}$ being the RIS-to-receiver(RIS$_i$-RX) channel \cite{RisASyed}. The action of the $M$ element RIS at the $i^{th}$ UE is represented as $\bold{\Phi}_i = \diag( e^{j\phi_1^i}, e^{j\phi_2^i}, \cdots, e^{j\phi_M^i})$. Here $\phi_n^i \in \Phi$, where $\Phi$ is a finite set phase angles with cardinality $K$. The dimensions of the precoder and combiner are $\bold{F}_D \in \mathbb{C}^{N_{rt} \times N}$, ${\bold{F}_A} \in \mathbb{C}^{N_t \times N_{rt}}$, ${\bold{W}_A^H} \in \mathbb{C}^{N_{rs} \times N_r}$, and ${\bold{W}_D^H} \in \mathbb{C}^{N \times N_{rs}}$.\\
The relationship between the transmitted signal vector $\bold{x}$ and the received symbol vector $\bold{y}$ at the receiver is given by
\begin{equation}\label{eq_2}
\begin{split}
\bold{y} = \bold{W}_D^H\bold{W}_A^H\bold{H}\bold{F}_A\bold{F}_D\bold{x} + \bold{W}_D^H\bold{W}_A^H \bold{n}. 
\end{split}
\end{equation}
Here, ${\bold{n}}$ is an $N_r\times1$ noise vector of independent and identically distributed (i.i.d) complex Gaussian random variables such that ${\bold{n}} \sim \mathcal{CN}(\bold{0},{\sigma_n^2}{\bold{I}_{N_r}})$. The precoders ${\bold{F}_D}$ and ${\bold{F}_A}$, and combiners ${\bold{W}_D^H}$ and ${\bold{W}_A^H}$ are designed for a given channel realization $\bold{H}$. We assume that the information about the channels $\bold{G}_i, \forall i$'s are known at the TX and the channel information $\bold{Q}_i, \forall i$'s are known at the RX. We further assume that the number of RF paths $N_{rs}$ on the receiver is the same as the number of parallel data streams $N$. The analysis is easy to extend and similar for the case $N_{rs} \ne N$ \cite{Zakir11}.

\subsection{MSE minimization problem formulation}\label{prob}
The MSE $\delta$ of the received and combined signal $\bold{y}$ using \eqref{eq_2} can be written as
\begin{equation}\label{eq_3}
\delta \triangleq \tr(\bold{M}(\bold{x})),
\end{equation}
where $\bold{M}(\bold{x})$ is the MSE matrix that can be written as
\begin{equation}\label{eq_4}
\begin{split}
\bold{M}(\bold{x}) &= (E\big[ (\bold{y}-\bold{x})(\bold{y}-\bold{x})^H\big]),\\
&=  p(\bold{K}-\bold{I}_{N})(\bold{K}-\bold{I}_{N})^H + \sigma_{n}^2\bold{W}\bold{W}^H,\text{ where}\\
\bold{K} &= \bold{W}_D^H \bold{W}_A^H \Big(\bold{H}_d + \sum_{i=0}^{N_d-1}\bold{Q}_i\bold{\Phi}_i\bold{G}_i \Big) \bold{F}_A\bold{F}_D.
\end{split}
\end{equation}
Here $E[{\bold{x}}{\bold{x}}^H] = p{\bold{I}_{N}}, \bold{W} = \bold{W}_D^H \bold{W}_A^H, E[{\bold{n}}{\bold{n}}^H] = {\sigma_n^2}{\bold{I}_{N_r}}$, and $p$ is the average power of the symbol $\bold{x}$.\\
\indent The design of the precoder, combiner, and the RIS phase-shift settings for all the $N_d$ cooperating UEs to minimize the MSE $\delta$ at the intended RX can be posed as a multi-dimensional optimization problem \cite{Zakir11}
%
\begin{small}
\begin{equation} \label{eq_5}
(\bold{F}_A^{opt},\bold{F}_D^{opt},{\bold{W}_A^H}^{opt},{\bold{W}_D^H}^{opt}, \{ \bold{\Phi}^{opt}_i \}_{i=1}^{N_d} ) = \hspace{-0.45in}
 \argminF_{\bold{F}_A,\bold{F}_D,{\bold{W}_A^H},{\bold{W}_D^H},\{ \bold{\Phi}_i \}_{i=1}^{N_d}} \hspace{-0.3in} \delta.
\end{equation}
\end{small}
Let $(\bold{F}_A^{no},\bold{F}_D^{no},{\bold{W}_A^H}^{no},{\bold{W}_D^H}^{no},\{ \bold{\Phi}_i^{no} \}_{i=1}^{N_d})$ be the near-optimal solution \footnote{\scriptsize A solution is said to be near-optimal if the MSE obtained using the same is $\delta^{no}$ satisfies the condition $|\delta^o - \delta^{no}| \le \epsilon$, where $\delta^o$ is the MSE due to the optimal solution and $\epsilon$ is a small number close to zero.} to the precoder, combiner, and the RIS phase settings of all the RIS elements of the cooperating devices. The MSE matrix in \eqref{eq_4} can now be written as
\begin{equation}\label{eq_6}
\begin{split}
\bold{M}(\bold{x}) &=  p(\bold{K}^{no}-\bold{I}_{N})(\bold{K}^{no}-\bold{I}_{N})^H + \sigma_{n}^2\bold{W}^{no}{\bold{W}^{no}}^H,\\
\text{ where }\bold{K}^{no} &=  \bold{W}^{no} \Big(\bold{H}_d + \sum_{i=0}^{N_d-1}\bold{Q}_i\bold{\Phi}_i^{no}\bold{G}_i \Big) \bold{F}_A^{no}\bold{F}_D^{no} \text{ and } \\
\bold{W}^{no} &= {\bold{W}_D^H}^{no} {\bold{W}_A^H}^{no}.
\end{split}
\end{equation}

\section{Physical layer procedures with UE-RIS and phase optimization}\label{Sol}
\begin{figure}[h!]
\centering
\includegraphics[scale=0.16]{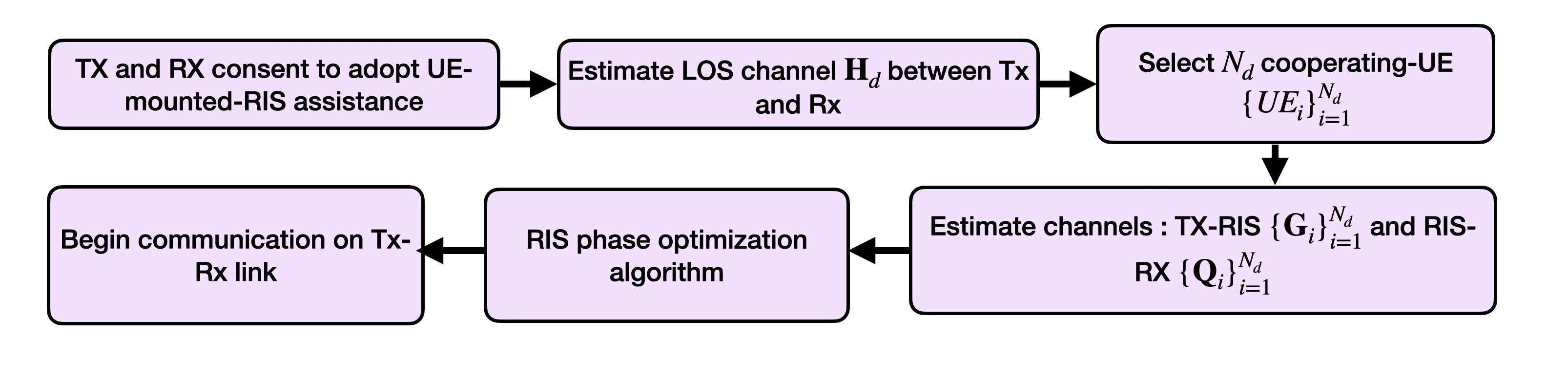}
\caption{\small PHY layer procedure to setup a UE-mounted RIS-assisted Tx-Rx communication}
\label{fig_phy}
\end{figure}

In this Section, we describe the physical layer procedures to establish a TX-RX communication by leveraging UE-mounted-RIS framework for near-optimal MSE performance. This is illustrated using the flow graph in Fig. \ref{fig_phy}.

\subsection{Cooperating-UEs selection}\label{ue_select}
\begin{figure}[h!]
\centering
\includegraphics[scale=0.14]{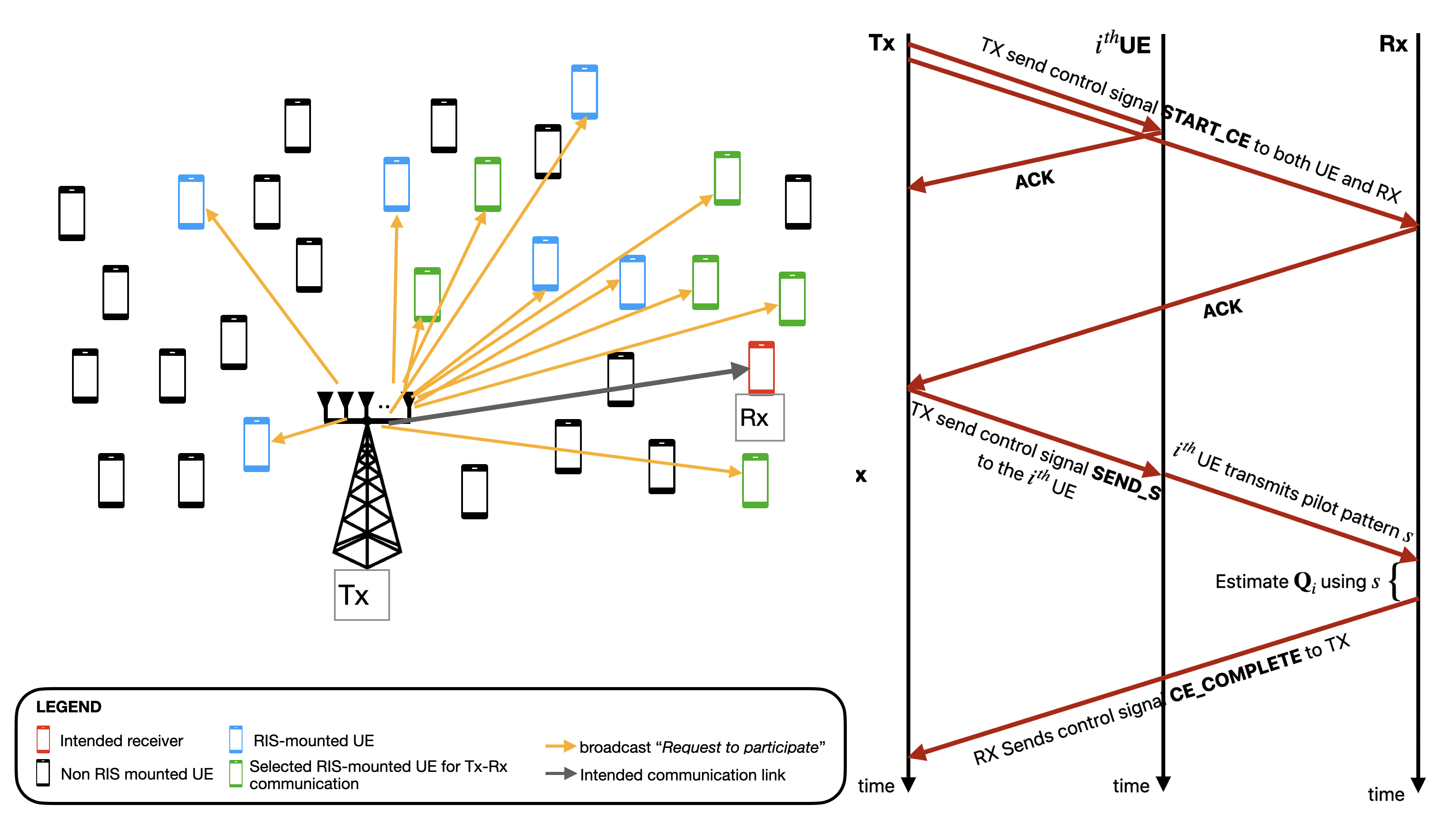}
\caption{\small Cooperating-UEs selection and channel estimation}
\label{fig_ue}
\end{figure}

Here, we describe an algorithm to identify the subset of RIS-mounted-UEs that are willing to participate and enable optimal TX-RX communication in the network. This algorithm is generally implemented at higher layers, although a tight physical layer dependency exists. The TX, typically a base station or an access point, sends a broadcast request to all the UEs that are RIS-equipped. Let $N_D$ be the number of UEs that respond to the TX. The TX then observes the estimates of the channels between the TX and the respective UEs, $\{ \bold{G}_i \}_{i=1}^{N_D}$ over a few coherence intervals to assess the mobility conditions (MC) of the UEs. The TX selects a subset of the UEs based on the low mobility scenarios. Let $L_M$ be the subset of them selected. The channel estimation (CE) is carried out to determine the channels $\{ \bold{Q}_i \}_{i=1}^{L_M}$ using Algorithm \ref{ChanEstAlgo}. Based on the observation of the channel quality (CQI) of $\{ \bold{Q}_i \}_{i=1}^{L_M}$ over a few coherence duration, the $N_d$ best units that will eventually participate in the UE-RIS framework are identified. The protocol is described briefly using Algorithm \ref{Algo_Select_UEs}, and illustrated using the Fig. \ref{fig_ue}.

\begin{algorithm}[h!]
  \caption{UE selection procedure}\label{Algo_Select_UEs}
  \begin{algorithmic}[1]
   \small
      \Procedure{UE Select}{}
      	 \State TX broadcast ``\textit{Request to participate}" to all RIS-UEs
	  \For{each UE $i$ that accepted the request}
	  	\State Tx performs the CE $\bold{G}_i$ between Tx and the UE $i$
		\State Using $\bold{G}_i$ TX evaluates CQ and MC	
          \EndFor
          \State TX identifies the best $L_M$ UEs based on CQ and low MC
          \State $\{ \bold{Q}_i \}_{i=1}^{N_D} \gets \textbf{ChanEst()}$ \Comment{ Evaluate UE-RX channels}
          \State TX identifies $N_d$ best UEs using $\{ \bold{Q}_i \}_{i=1}^{N_D}$
	 \State{\textbf{return} $\{ UE \}_{i=1}^{N_d}$}\Comment{ Return the identified RIS-UEs}
  \EndProcedure
  \end{algorithmic}
\end{algorithm}

\subsection{Channel estimation}\label{chan_est1}
\begin{figure}[b!]
\centering
\includegraphics[scale=0.18]{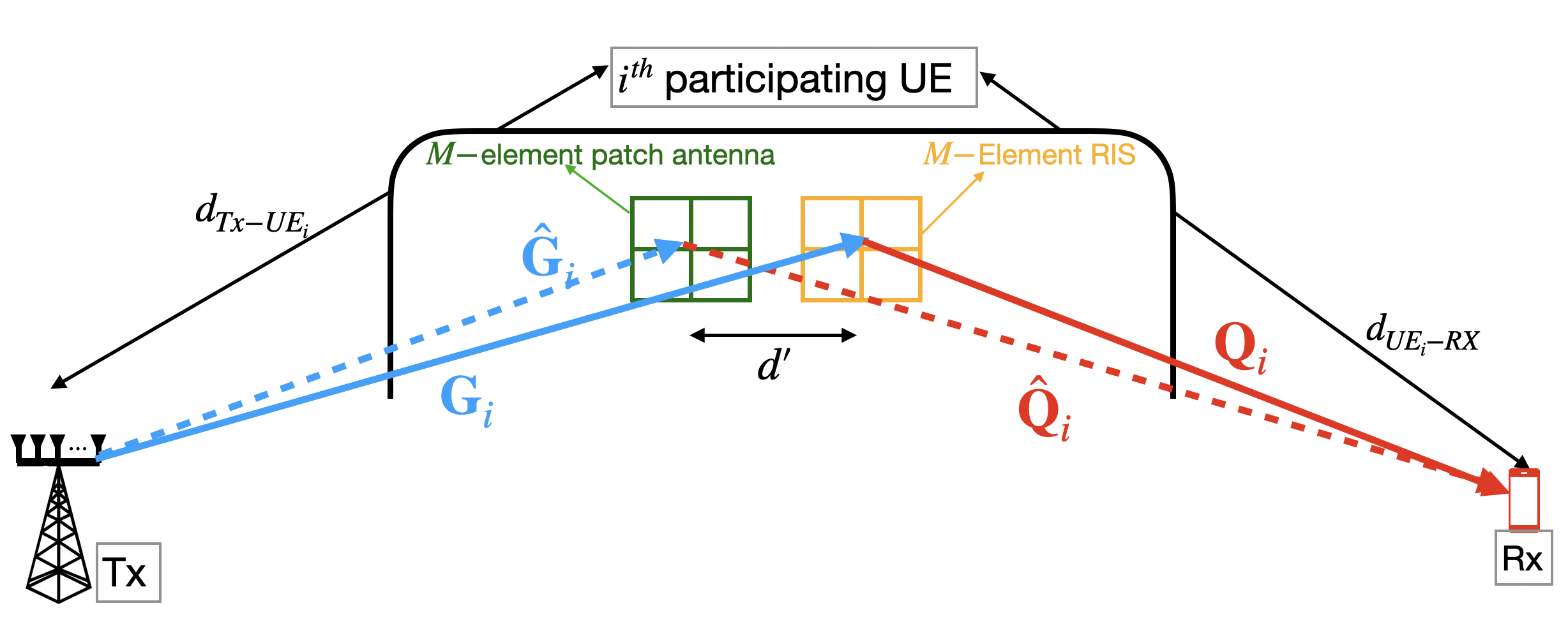}
\caption{\small UE-RIS channel estimation}
\label{fig_ce}
\end{figure}
In this subsection, we discuss the channel estimation of the channels $\{ \bold{G}_i \}_{i=1}^{N_d}$, $\{ \bold{Q}_i \}_{i=1}^{N_d}$, and $\bold{H}_d$. Let $\bold{\hat{G}}_i$ be the effective channel between the TX and the antennas of the 
$i^{\text{th}}$ UE. Similarly, let $\bold{\hat{Q}}_i$ be the effective channel between the $i^{\text{th}}$ UE and the RX. Also, let $d_{TX-UE_i}$ and $d_{UE_i-RX}$ be the distance between the TX and the antenna patch on the $i^{\text{th}}$ UE and distance between the antenna patch on the $i^{\text{th}}$ UE to the RX, respectively, as illustrated in the Fig. \ref{fig_ce}. If $d'$ be the effective distance between RIS elements to the antenna patch on the $i^{\text{th}}$ UE, and since $d_{Tx-UE_i} \gg d'$, and $d_{UE_i-RX} \gg d'$, we can write
\begin{equation}
\begin{split}
\lVert \hat{\bold{G}_i} - \bold{G}_i \rVert_F^2 & \rightarrow 0,\\
\lVert \hat{\bold{Q}_i} - \bold{Q}_i \rVert_F^2 & \rightarrow 0.
\end{split}
\end{equation}
Hence it suffices to estimate the channels $\hat{\bold{G}_i}$ and $\hat{\bold{Q}_i}$ for all $i \in [1, N_d]$. That is $\bold{G}_i \rightarrow \hat{\bold{G}_i}$ and $\bold{Q}_i \rightarrow \hat{\bold{Q}_i}$. Thus the channel estimation of $\bold{G}_i$ and $\bold{Q}_i$ turns out to be a traditional pilot-based method owing to the RIS being mounted on an active computing and communicating device.\\ 
\indent The channels $\hat{\bold{G}_i}$ is available with both the TX as well as the $i^{\text{th}}$ participating UE as part of the communication link setup procedure between the TX and the cooperating UE $i$. Hence, the channels $\{ \bold{G}_i \}_{i=1}^{N_d}$ are known to the TX as well as the corresponding UEs. The direct LOS channel $\bold{H}_d$, if available, is estimated in the same fashion as above by the RX using the pilot symbols transmitted from the TX. Similarly, the channels between the RX and the cooperating UEs with RIS are carried out asynchronously by using a specific pilot pattern $s$ known to all the participating UEs, TX, and RX. A possible candidate algorithm is elucidated in Algorithm \ref{ChanEstAlgo}. 
\begin{algorithm}[t!]
  \caption{Channel Estimation for $\{\bold{Q}_i\}_{i=1}^{N_d}$}\label{ChanEstAlgo}
  \begin{algorithmic}[1]
   \small
      \Procedure{ChanEst($s$)}{}
         \State The pilot pattern $s$ is known to $\{\text{UE}_i\}_{i=1}^{N_d}$, TX, and RX
         \For{each participating UE $i$}
                 \State TX sends control signal START-CE to UE $i$ and RX
                 \State TX waits for ACK from both UE $i$ and RX 
		\State TX sends control signal SEND-S to UE $i$
		\State RX performs channel estimation using $s$ to get $\bold{Q}_i$
		\State TX waits for CE-COMPLETE from RX		
          \EndFor
         \State{\textbf{return} $\{ \bold{Q}_i \}_{i=1}^{N_d}$}
  \EndProcedure
  \end{algorithmic}
\end{algorithm}
An illustration of the channel estimation procedure for $\bold{Q}_i$ for the $i^{th}$ participating UE is described using a timing diagram in Fig. \ref{fig_phy}. The sequence of operations is repeated for all the participating UEs sequentially.

\subsection{RIS phase optimization algorithm}\label{Algo}
Joint optimization of all the RIS elements on the participating UEs in the UE-RIS framework with a large number of UEs ensures that the effective channel in \eqref{eq_1} is well conditioned such that the entries of the channel entries $[ \bold{H} ]_{i,j} \sim \mathcal{CN}(0,{\sigma_n^2})$. This enables higher spatial multiplexing gain with improved MSE and capacity \cite{Vishwa}. However, since the effective channel $\bold{H}$ is a function of the RIS phase shifts \eqref{eq_1}, the design of the precoder and combiner at TX and RX is not straightforward. We propose an alternating optimization (AO) algorithm to jointly optimize the hybrid precoder and combiner $(\bold{F}_A,\bold{F}_D,{\bold{W}_A^H},{\bold{W}_D^H})$, and the RIS phase settings of the cooperating UEs $\{ \bold{\Phi}_ i\}_{i=1}^{N_d}$ alternately that solves the optimization problem \eqref{eq_5}.
\subsubsection{Precoder and combiner design}\label{prec_goa}
The hybrid precoding and combing techniques for systems employing phase shifters in mmWave/Thz transceiver architectures impose constraints on them. The analog precoder $\bold{F}_A$ and combiner ${\bold{W}_A^H}$ entries need to satisfy unit norm entries in them \cite{SigProc,PreDsgn,Zakir7,Zakir11}. The analog precoder $\bold{F}_A^j$ and the digital precoder $\bold{F}_D^j$ at the $j^{\text{th}}$ iteration of the AO algorithm are arrived such that
\begin{equation}\label{eq_11}
{\bold{V}^j}^H \bold{F}_A^j \bold{F}_D^j \approx \bold{I}_M,
\end{equation}
where ${\bold{V}^j}^H$ is a right unitary matrix of the singular value decomposition (SVD) of the effective channel $\bold{H}^j$ at the $j^{\text{th}}$ iteration of the AO algorithm. That is, $\bold{H}^j = \bold{U}^j\bold{\Sigma}^j{\bold{V}^j}^H$ with the matrix $\bold{U}^j$ and $\bold{\Sigma}^j$ being the left unitary matrix and singular value matrix of the SVD decomposition at $j^{\text{th}}$ iteration, respectively.\\
The hybrid precoders are derived upon solving the optimization problem \cite{PreDsgn,SigProc} stated below. 
\begin{equation}\label{eq_12}
\begin{aligned}
({\bold{F}_A^j}^{no},{\bold{F}_D^j}^{no}) = & \argminF_{{\bold{F}_D^j},{\bold{F}_A^j}}{\lVert {\bold{V}^j - {{\bold{F}_A^j}{\bold{F}_D^j}}} \rVert }_F, \\
\text{such that } & {\bold{F}_A^j}\in{\mathcal{F}_{RF}}, {\lVert {{\bold{F}_A^j}\bold{F}_D^j} \rVert }_F^2 = N.
\end{aligned}
\end{equation}
The set $\mathcal{F}_{RF}$ is the set of all possible analog precoders that correspond to a hybrid precoder architecture based on phase shifters. This includes all possible $N_t \times N_{rt}$ matrices with constant magnitude entries.\\
\indent Similarly, the analog combiner ${\bold{W}_A^j}^H$ and the digital combiner ${\bold{W}_D^j}^H$ at the $j^{\text{th}}$ iteration of the AO algorithm are designed such that
\begin{equation}\label{eq_13}
\begin{split}
{\bold{W}_D^j}^H\ {\bold{W}_A^j}^H\bold{U}^j \approx \bold{I}_M.
\end{split}
\end{equation}
The hybrid combiners are derived using \cite{PreDsgn}
\begin{equation}\nonumber
\begin{split}
({{\bold{W}_A^j}^H}^{no},{{\bold{W}_D^j}^H}^{no}) &= \argminF_{{{\bold{W}_D^j}^H},{{\bold{W}_A^j}^H}}{\lVert {{\bold{U}^j}^H - {{\bold{W}_D^j}^H}{{{\bold{W}_A^j}^H}}} \rVert }_F, \\
\end{split}
\end{equation}
\begin{equation}\label{eq_14}
\text{such that }{{\bold{W}_A^j}^H}\in{\mathcal{W}_{RF}}, {\lVert {{{\bold{W}_D^j}^H}{{\bold{W}_A^j}^H}} \rVert }_F^2 = N.
\end{equation}
Here again, the set $\mathcal{W}_{RF}$ is the set of all possible analog combiners that correspond to hybrid combiner architecture based on phase shifters. This includes all possible $N_{rs} \times N_r$ matrices with constant magnitude entries.
\subsubsection{RIS phase shift optimization in AO}\label{ris_goa}
\indent Next, the RIS phase shift solutions $\{ \bold{\Phi}_ i^j\}_{i=1}^{N_d}$ at the $j^{\text{th}}$ iteration is obtained by solving the optimization problem \eqref{eq_15} derived from \eqref{eq_5}, \eqref{eq_6}.
\begin{equation}\label{eq_15}
\begin{split}
\{ \bold{\hat{\Phi}}_ i^j\}_{i=1}^{N_d} &= \argminF_{\{ \bold{\Phi}_ i\}_{i=1}^{N_d}}{\tr \{ \bold{M}^j(\bold{x}) \}},\text{ where}\\
\tr \{ \bold{M}^j(\bold{x}) \} &= \mathcal{L}({\bold{F}_A^j}^{no},{\bold{F}_D^j}^{no},\{ \bold{\Phi_i}^j \}_{i=1}^{N_d},{{\bold{W}_A^j}^H}^{no},{{\bold{W}_D^j}^H}^{no}),\\
&= \tr \{  p(\bold{K}^j-\bold{I}_{N})(\bold{K}^j-\bold{I}_{N})^H + \sigma_{n}^2\bold{W}^j{\bold{W}^j}^H \}.
\end{split}
\end{equation}
The RIS optimization in \eqref{eq_15} can be solved \textit{optimally} using exhaustive search (AO-ES). However, AO-ES becomes prohibitively expensive computationally for practical implementation when $N_d$ or $M$ is large. We propose to use IDBP algorithm for RIS phase optimization in the AO framework (AO-IDBP) \cite{Zakir11}. The IDBP guarantees near-optimal solution with significant computational advantage.
\begin{breakablealgorithm}
  \caption{AO-IDBP}\label{Algo_AO}
  \begin{algorithmic}[1]
   \small
      \Procedure{AO}{$\bold{H}_d, \{ \bold{G}_i \}_{i=1}^{N_d},\{ \bold{Q}_i \}_{i=1}^{N_d},\epsilon_T$}
                 \State{$\text{Solve }\{ \bold{\Phi}_ i^0\}_{i=1}^{N_d} \gets $initial values using IDBP algorithm \cite{Zakir11}}
		\State{$\bold{H}^0 \gets \bold{H}_d + \sum_{i=0}^{N_d-1} \bold{Q}_i\bold{\Phi}_i^0\bold{G}_i$}
                	\State{$(\bold{U}^0,\bold{\Sigma}^0,{\bold{V}^0}^H) \gets \text{SVD}(\bold{H}^0)$}
                	\State{$\text{Evaluate }(\bold{F}_A^0,\bold{\tilde{F}}_D^0)\text{ using \eqref{eq_12}}$}	
	        \State{$\text{Evaluate }({\bold{W}_A^0}^H, {\bold{\tilde{W}}_D}^{0^H} )\text{ using \eqref{eq_14}}$}
	        \State{$\delta_0 \gets \mathcal{L}(\bold{F}_A^{0},\bold{\tilde{F}}_D^{0},\{ \bold{\Phi_i}^0 \}_{i=1}^{N_d},{\bold{W}_A^H}^{0},\bold{\tilde{W}}_D^{H^0})$}
		\State{$k \gets 1$}
      		\Do
			\State{$\text{Solve }\{ \bold{\Phi}_ i^k\}_{i=1}^{N_d}\text{ in \eqref{eq_15} using IDBP \cite{Zakir11}}$}
			\State{$\bold{H}^k \gets \bold{H}_d + \sum_{i=0}^{N_d-1} \bold{Q}_i\bold{\Phi}_i^k\bold{G}_i$}
                		\State{$(\bold{U}^k,\bold{\Sigma}^k,\bold{V}^{k^H}) \gets \text{SVD}(\bold{H}^k)$}
                		\State{$\text{Evaluate }(\bold{F}_A^k,\bold{\tilde{F}}_D^k)\text{ using \eqref{eq_12}}$}	
	                 \State{$\text{Evaluate }({\bold{W}_A^k}^H, \bold{\tilde{W}}_D^{k^H})\text{ using \eqref{eq_14}}$}   
	                 \State{$\delta_{k} \gets \mathcal{L}(\bold{F}_A^{k},\bold{\tilde{F}}_D^{k},\{ \bold{\Phi_i}^k \}_{i=1}^{N_d},{\bold{W}_A^H}^{k},\bold{\tilde{W}}_D^{H^k})$}
	                 \State{$err \gets \delta_{k-1} - \delta_{k}$}  
	             	\State{$k \gets k + 1$}
		\doWhile{$\{ err > \epsilon_T \}$} 
  \EndProcedure
\end{algorithmic}
\end{breakablealgorithm}

\section{Simulations}\label{sim}
\begin{figure}[h!]
\begin{center}
\includegraphics[width=1.0\linewidth]{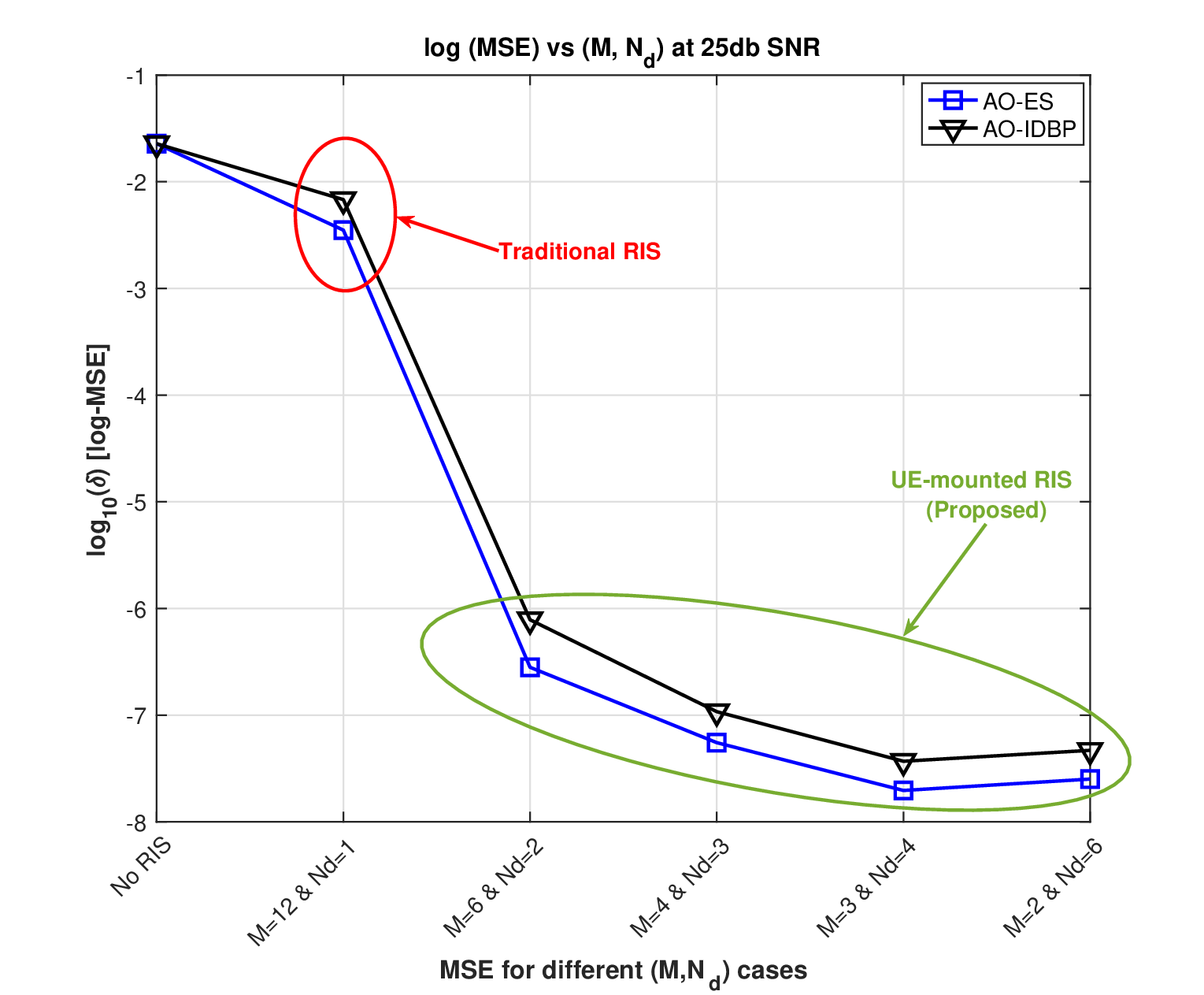}
\end{center}
\caption{\scriptsize Simulation results}\label{fig2}
\end{figure}

To evaluate the proposed framework with AO-IDBP algorithm, we run the simulations that compare the no-RIS case, a traditional RIS system with $M=12$ reflecting elements on a single UE, and a few other configurations such that $N_d \times M = 12$. To establish a benchmark, we also evaluate the above cases using the Exhaustive Search (ES) algorithm for RIS phase-shift identification in the AO framework (AO-ES). The ES always guarantees an optimal solution to RIS phase shifts at every iteration of AO-ES. The following parameter values are used in our simulations- $N_t = 12, N_r=16, N=8, K=3$, TX-RX separation of $60$ meters, carrier frequency of $28$Ghz, baseband signal composed of 200 symbols of size $N \times 1$ mapped unto a $64-$QAM constellation, and a receive SNR=$25$dB. The key takeaway from the simulation results in Fig. \ref{fig2} is the fact that a large number of participating devices having fewer or even a single RIS element on them have excellent performance compared to a traditional single monolithic RIS that has a large number of reflecting elements $M$ thus substantiating the main claim of our work.\\

\section{Conclusion}\label{conc}
In conclusion, the future generations of wireless communication envision a MaMIMO-enabled dense network with a large number of available UEs operating in either millimeter wave or terahertz frequency ranges. Many UEs in such dense networks are rarely in motion in many practical scenarios. The proposed UE-mounted RIS system finds a sweet spot in such paradigms, boosting the performance of communication between Tx and Rx that experience poor channel conditions.

\bibliographystyle{IEEEtran}
\bibliography{ITWC_CRIS_BibTexFile}

\end{document}